\documentclass[twocolumn,secnumarabic,amssymb, nobibnotes, aps, prd]{revtex4-2}
 
\usepackage[american]{babel} 
\usepackage{graphicx}
\usepackage{dcolumn}
\usepackage{bm}
\usepackage{hyphenat}
\usepackage{amssymb}   
\usepackage{verbatim}
\usepackage{xpatch}
\usepackage{color}
\usepackage{epstopdf}
\usepackage{multirow}
\usepackage{mhchem} 
\usepackage{slashbox}

\usepackage{color, colortbl}
\definecolor{Gray}{gray}{0.9}

\usepackage{hyperref}

\begin{document}

\title{Periodic table for highly charged ions}

\author{Chunhai Lyu}
\altaffiliation{chunhai.lyu@mpi-hd.mpg.de}
\affiliation{Max-Planck-Institut f\"{u}r Kernphysik, Saupfercheckweg 1, 69117 Heidelberg, Germany}
\author{Christoph H. Keitel}
\affiliation{Max-Planck-Institut f\"{u}r Kernphysik, Saupfercheckweg 1, 69117 Heidelberg, Germany}
\author{Zolt\'{a}n Harman}
\affiliation{Max-Planck-Institut f\"{u}r Kernphysik, Saupfercheckweg 1, 69117 Heidelberg, Germany}

\date{\today}

\begin{abstract}

Mendeleev's periodic table successfully groups atomic elements according to their chemical and spectroscopic properties. However, it becomes less sufficient in describing the electronic properties of highly charged ions (HCIs) in which many of the outermost electrons are ionized. In this work, we put forward a periodic table particularly suitable for HCIs. It is constructed purely based on the successive electron occupation of relativistic orbitals. While providing a much-simplified description of the level structure of highly charged isoelectronic ions -- essential for laboratory and astrophysical plasma spectroscopies, such a periodic table predicts a large family of highly forbidden transitions suitable for the development of next-generation optical atomic clocks.
Furthermore, we also identify universal linear $Z$ scaling laws ($Z$ is the nuclear charge) in the so-called ``Coulomb splittings'' between angular momentum multiplets along isoelectronic sequences, complementing the physics of electron-electron interactions in multielectron atomic systems.   

\end{abstract}

\maketitle

%

\textit{Introduction} ---  
In multielectron atomic physics, the `building-up' principle~\cite{foot2005atomic} states that the electron would fill atomic orbitals according to their relative energies. For neutral atoms, the repulsion interaction between electrons renders the $(n+1)s,(n+1)p$ orbitals asserting energies lower than those of the $nd,nf$ orbitals, with $n$ being the principal quantum number, and $s,p,d,f$ corresponding to orbital angular momenta $l=0,1,2,3$, respectively. As a consequence, the electron would fill atomic orbitals according to the increasing value of $n+l$. This principle, also called Madelung--Janet rule~\cite{foot2005atomic}, determines the ground-state electron configurations, thus the physical and chemical properties of the corresponding transition metal, lanthanoid and actinoid elements in the conventional periodic table~\cite{schwerdtfeger2020periodic}. 

However, when the outer-shell electrons of an atom are stripped off, it becomes a highly charged ion (HCI). Though such systems are exotic, they play a key role in laboratory and astrophysical plasmas~\cite{beiersdorfer2003laboratory,Fe3C3D-2012,schmoger2015coulomb}, x-ray lasers~\cite{zhang1997saturated,depresseux2015table,lyu2020narrow}, nanolithography~\cite{Versolato2024}, tumor therapy~\cite{RevModPhys.82.383}, optical clocks~\cite{HCIclock-Ar13-2022,ludlow2015optical,kozlov2018highly}, and in testing  fundamental theories~\cite{morgner2023stringent,U90-2024,HCIclock-4f5s-2010}. As the electrons in HCIs feel a much stronger Coulomb interaction with the nucleus, the electrons would follow the so-called Coulomb filling rule~\cite{goudsmit1964order}: i.e., before occupying the $(n+1)$-th shell, the electron will first fill out all the orbitals in the $n$-th shell.

\begin{figure*}[t!]
\includegraphics[width=0.95\textwidth]{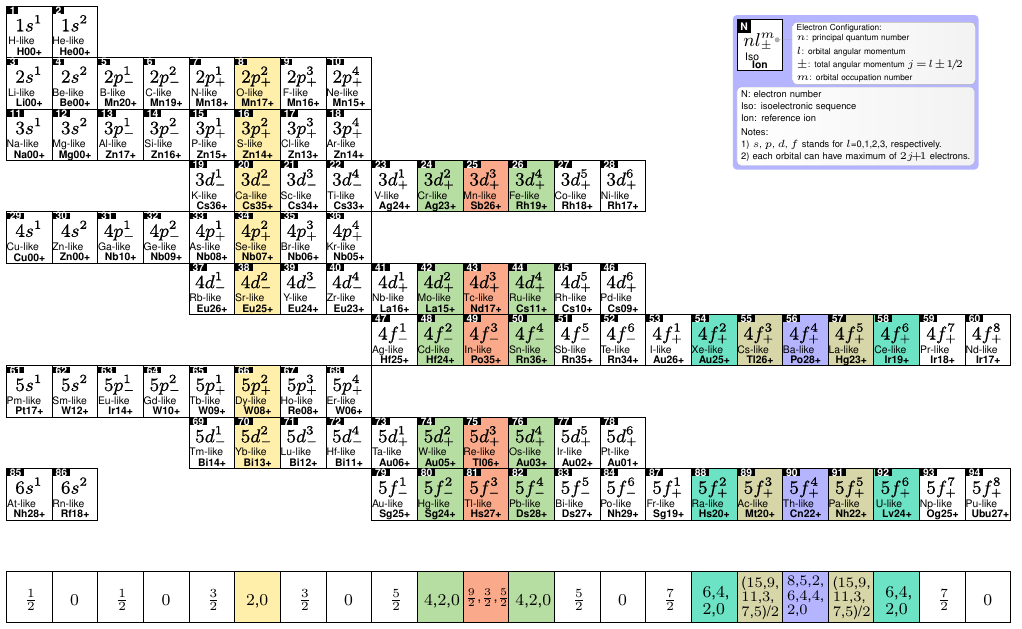}
\caption{\label{pd}\textbf{Periodic table for HCIs.} It is arranged according to the number of electrons $N$, with each cell represents one isoelectronic sequence with a given $nl_{{}^{\pm}}^m$ configuration. The reference ions at the bottom of each cell indicate the elements in which $jj$ couplings become significant. As the energy of $6s$ is always lower than that of $5f_{{}^+}$, the two cells with $N=85,86$ at the bottom-left corner of the table shall be inserted between the $5f_{{}^-}^6$ ($N=84$) and $5f_{{}^+}^1$ ($N=87$) cells. The bottom row presents all allowed $J$s, ordered according to their increasing energies, of the ground-state $j^m$ multiplet in each column. 
This periodic table is created via Latex by modifying the pgf-PeriodicTable package~\cite{pgf2}.} 
\end{figure*}

\textit{Periodic table} ---  
Here, we show that the Coulomb filling rule renders it possible to construct a meaningful new periodic table for HCIs. This allows us to gain extraordinary insight into the physics and applications of such strong-field multielectron systems. As shown in Fig.~\ref{pd}, in the newly proposed periodic table, each cell represents an isoelectronic sequence, containing tens of ions sharing the same relativistic valence-electron configuration $nl_{{}^\pm}^m$ ($m$ is the number of electrons in the corresponding relativistic orbital, and the subscripts $\pm$ stand for the single-electron total angular momentum $j\!=\!l\!\pm\!1\!/\!2$, respectively). Then, each column groups cells with the same $j^m$ multiplet, and each row (period) collects cells with the same $nl_{{}^\pm}$ values, where $ns$, $np_{{}^-}$ and $np_{{}^+}$ orbitals are put in the same row to save space. As both $np_{{}^+}$ and $nd_{{}^-}$ have $j\!=\!3\!/\!2$, and both $nd_{{}^+}$ and $nf_{{}^-}$ have $j\!=\!5\!/\!2$, they are grouped into the same column, respectively. 

Different from the conventional periodic table where $nl_{{}^-}$ and $nl_{{}^+}$ orbitals are treated equivalently due to strong $LS$ couplings, in our case for HCIs with strong $jj$ couplings, the electrons will first fill the $nl_{{}^-}$ orbital. Only after $nl_{{}^-}$ is fully occupied with $2j+1$ electrons, the remaining electron can start to fill the $nl_{{}^+}$ orbital. Such an arrangement is straightforward as Dirac's relativistic theory~\cite{grant2007relativistic,dirac1981principles} predicts a lower orbital energy for $nl_{{}^-}$ than that for its $nl_{{}^+}$ counterpart. However, to have the abovementioned filling order, the fine-structure splittings between $nl_{{}^-}$ and $nl_{{}^+}$ shall exceed the energy caused by electron-electron interactions. The ion at the bottom of each cell benchmarks the lightest element at which this rule is fulfilled, thus resembles the crossing point from $LS$ couplings to $jj$ couplings.

This new periodic table would bear significant consequences for the understanding and applications of HCIs in fields ranging from atomic physics and quantum optics, to plasma spectroscopy and astrophysics, and to ultraprecise clocks.

\textit{Ground states} ---
As a first consequence, the periodic table in Fig.~\ref{pd} enormously simplifies the spectroscopic analysis of complex atomic structures. It is well known that atomic systems with partially filled $d$ and $f$ orbitals contain tremendously complicated fine-structure terms~\cite{grant2007relativistic,atkins2023atkins}. Without numerical calculations, it is difficult to identify their ground states~\cite{rodrigues2004systematic}. However, our new periodic table introduces additional closed-shell configurations, namely, the columns with fully occupied $np_{{}^-}^2$, $nd_{{}^-}^4$, and $nf_{{}^-}^6$ configurations. Based on this, it allows one to define new valence electron configurations and their corresponding hole states. Therefore, the ground state of HCIs can be easily identified by analyzing the angular couplings of these simplified electron configurations. The results are shown at the bottom of the periodic table in Fig.~\ref{pd}, where all possible multielectron total angular momenta $J$ are listed for each column.     

Take the open-shell $nd^5$ ions for instance, they contain 37 fine-structure terms with $J\!=\!1\!/\!2$ up to $13\!/\!2$~\cite{fischer2019grasp2018}, and it is not obvious which term has the lowest energy.  
However, in our periodic table, this corresponds to an electron configuration of $[nd_{{}^-}^4]nd_{{}^+}^1$, with the $nd_{{}^-}$ orbital being fully occupied. The single-valence $nd_{{}^+}^1$ electron immediately indicates that $J\!=\!5\!/\!2$ is the only term, and thus the ground state of the corresponding HCIs. In the case of $4f^7$ ions containing 327 fine-structure states~\cite{fischer2019grasp2018}, the new notation $[4f_{{}^-}^6]4f_{{}^+}^1$ directly indicates that the ground state comes with $J\!=\!7\!/\!2$. For systems with 2, 3, and 4 valence electrons in a given $nl_{{}^{\pm}}$ orbital, the Pauli's exclusion principle restricts the possible $jj$-coupling schemes, thus greatly reduces the number of states in the ground-state configurations. At the bottom row of the periodic table Fig.~\ref{pd}, the corresponding $J$s are listed in ascending order based on their energies. As expected by Hund's rule~\cite{foot2005atomic}, the state of the highest $J$ is the ground state. Furthermore, by sharing the same $J$ values, it is apparent that the $nd_{{}^+}^1$ and $nd_{{}^+}^2$ configurations are symmetric with their hole states $nd_{{}^+}^5$ and $nd_{{}^+}^4$, respectively, a phenomenon that also applies to the $f_{{}^-}$ and $f_{{}^+}$ ions. 

\begin{figure*}[t!]
\includegraphics[width=0.95\textwidth]{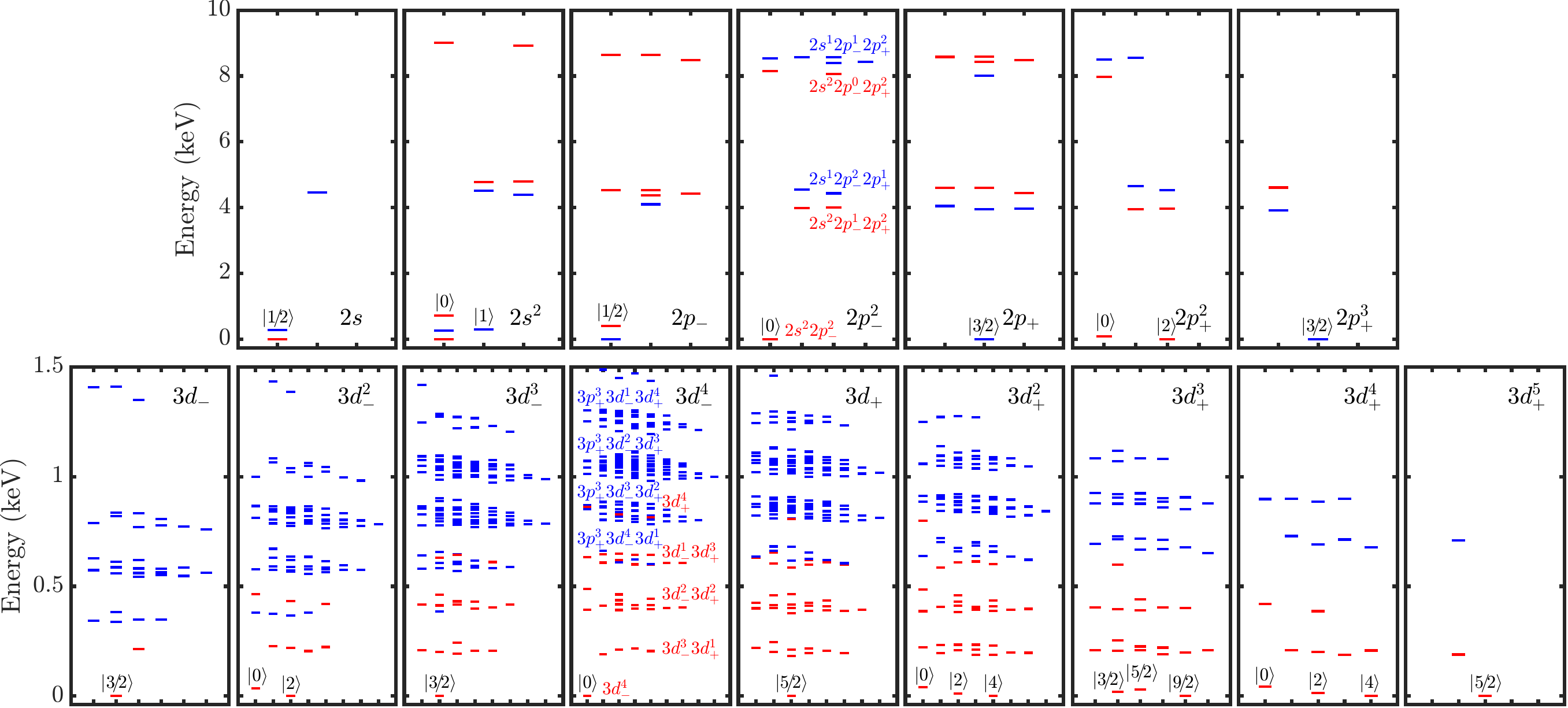}
\caption{\label{levels}\textbf{Clustered level structures in highly charged uranium ions:} the upper and lower panels correspond to the ions from the second and fourth rows of the periodic table shown in Fig.~\ref{pd}, respectively. Horizontal axis is the total angular momentum, with the first tick point being $0$ (or $1\!/\!2$) for integer (or half-integer) $J$s. The parity is colored in red (blue) for even (odd) states. In the lower panel, levels in blue color are the states formed by single excitation mainly from the $3p_{{}^+}$ subshell, with the level energy around 1.4~keV being the single excitation from the $3p_{{}^-}$ subshell. } 
\end{figure*}

\textit{Excited states} --- 
As a second consequence, the relativistic notation of the electronic configurations can reveal electron transitions between the $l_{{}^-}$ and $l_{{}^+}$ subshells, and thus 
exposing the grouped low-lying level structures of HCIs under strong $jj$ couplings. To demonstrate this, the excited states of uranium ions from the second ($2s,~2p_{{}^-}$ and $2p_{{}^+}$ ions) and the fourth ($3d_{{}^-}$ and $3d_{{}^+}$ ions) rows of the periodic table, calculated with the GRASP2018 codes~\cite{fischer2019grasp2018}, are shown in Fig.~\ref{levels}. 
From the upper panel, a distinct grouping of the $2s$ and $2p_{{}^-}$ orbitals and their sequential excitations to the high-lying $2p_{{}^+}$ orbitals are observed. The better proximity of the $2p_{{}^-}$ to the $2s$ than the $2p_{{}^+}$ orbital, resembling a screened hydrogenlike system, is a direct consequence of strong relativistic effects in HCIs.

In the lower panel of Fig.~\ref{levels}, the complicated fine-structure levels (red color) of each $3d_{{}^\pm}^m$ ion are clustered into 1, 2, 3 and 4-electron excitations from the $3d_{{}^-}$ to the $3d_{{}^+}$ orbital. The similarities in the level structures between the $3d_{{}^-}^{4-q}3d_{{}^+}^q$ and $3d_{{}^-}^{q}3d_{{}^+}^{6-q}$ excited configurations ($q\leq4$ is the number of excitations) recover a particle-hole symmetry in both $3d_{{}^-}$ and $3d_{{}^+}$ subshells.   
Limited by dense, complex atomic states in multiply excited ions, state-of-the-art collisional-radiative atomic plasma models~\cite{chung2005flychk,rosmej2021plasma} usually employ superconfigurations~\cite{chung2005flychk} that mainly account for transitions between principal orbitals, to simulate ionic populations and fluorescence in plasmas. The grouped level structures observed in Fig.~\ref{levels} bring forth a manageable approach to extend these models to include transitions between states based on relativistic configurations~\cite{kaastra2024_spex}. This would allow us to resolve the multielectron excitations and ionizations of relativistic subshells, and to gain novel insight into the astrophysical and laboratory atomic plasma processes in fields ranging from laser-produced plasma~\cite{depresseux2015table,lyu2020narrow} and warm dense matter~\cite{mercadier2024transient}, to confined fusion~\cite{HCI-W-2008,HCI-W-2009,HCI-fusion-2012} and to solar systems~\cite{Fe3C3D-2012}. 

Furthermore, the lifetimes of these excited configurations, manifested by magnetic-dipole decays, are in the range of 10~nanoseconds. This corresponds to a time scale similar to that in the excited states of neutral atoms manifested by electric-dipole allowed decay in the optical range~\cite{scully1997quantum,brif2010control}. Thus, the abundant low-lying $\varLambda$-, $\varXi$-, and $V$-type level structures observed in  Fig.~\ref{levels} are ideal experimental platforms for the development of x-ray quantum optics~\cite{adams2013x,heeg2021coherent,heeg2017spectral,haber2017rabi,cavaletto2014broadband,EUVclock-lyu}.

\textit{Forbidden transitions} --- 
The third consequence from this new periodic table is that it predicts a large  family of highly forbidden optical and XUV transitions. This may enable the construction of next-generation atomic clocks~\cite{ludlow2015optical,4f12ion-2012,lyu2025ultrastable,Kathrin2023}. The corresponding transitions are from ions highlighted by the colored columns of the periodic table shown in Fig.~\ref{pd}, where the valence orbitals are multiply occupied. Due to the Pauli exclusion principle in orbital filling, the ground-state configurations of these ions contain multiple $J$s whose values differ from each other by 2 or 3 (see the button row of Fig.~\ref{pd}). As a result, the lifetimes of the corresponding excited states are dominated by decays via highly forbidden electric quadrupole ($E2$) or magnetic octupole ($M3$) transitions that allow for the development of ultrastable HCI clocks~\cite{HCIclock-Ar13-2022}. To be more explicit, the level structures of the $2p_{{}^+}^2$, $3d_{{}^-}^2$, $3d_{{}^+}^2$, $3d_{{}^+}^3$ and $3d_{{}^+}^4$ ions can be found in Fig.~\ref{levels}. Within these ground-state configurations, the low-lying levels are well isolated from the states involving excitations from the $l_{{}^-}$ to the $l_{{}^+}$ subshells. 
Take the $3d_{{}^+}^2$ and $3d_{{}^+}^4$ configurations for example, they are the 2-electron and 2-hole states of the $3d_{{}^+}$ orbital. Therefore, within the ground-state configurations, they share the same level structure: i.e., with $\left|4\right\rangle$ being the ground state, the $\left|2\right\rangle$ state is the first excited state whose $100\!\sim\!1000$-s lifetime is manifested by slow $E2$ decay. Thus, the $\left|4\right\rangle\!\rightarrow\!\left|2\right\rangle$ transition in both systems can serve as an ultrastable clock reference with projected instabilities~\cite{ludlow2015optical} $\sigma_y\sim10^{-17}/\sqrt{\tau}$ ($\tau$ is the average measuring time).

\begin{figure}[t!]
\includegraphics[width=0.54\textwidth]{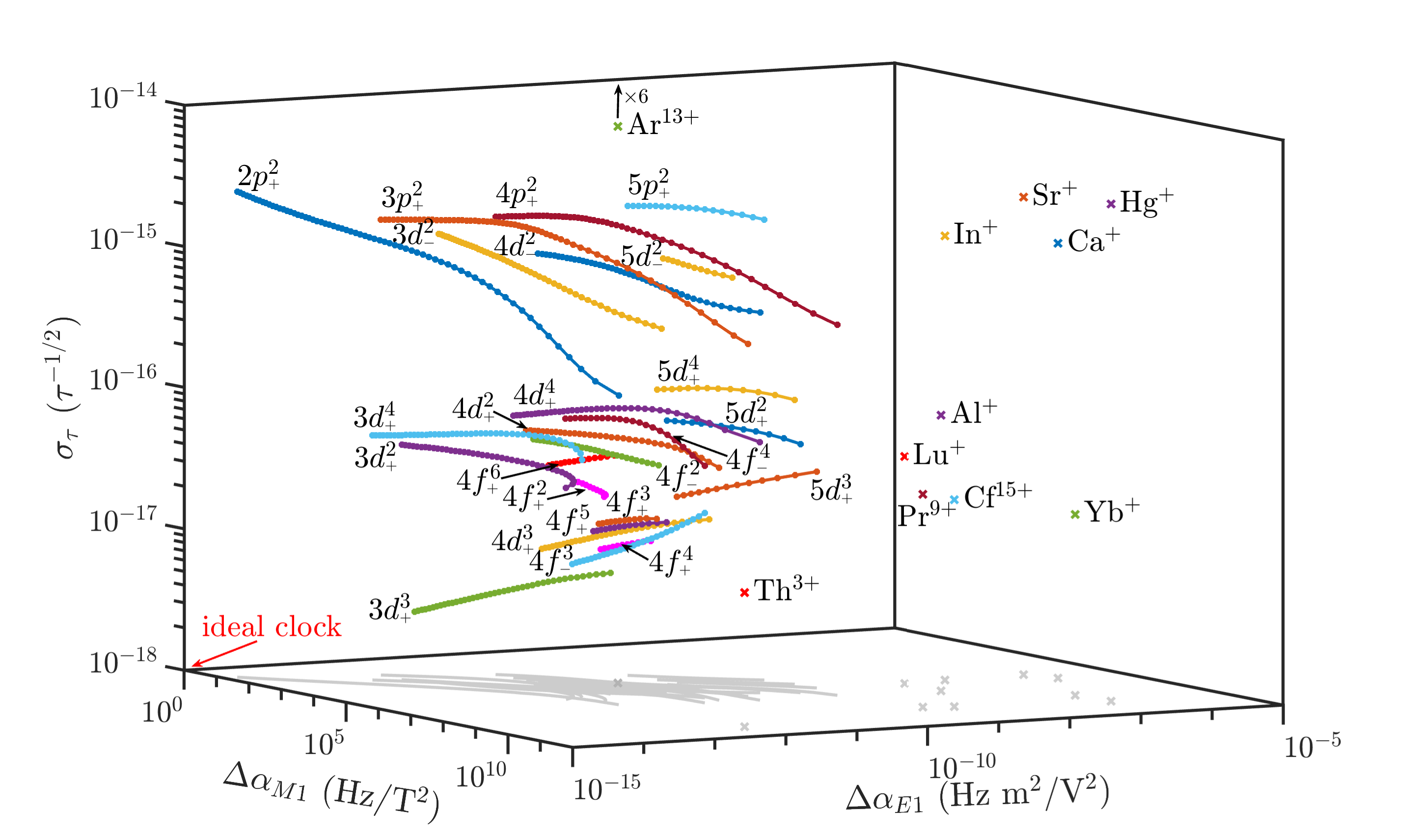}   
\caption{\label{clock}\textbf{Superior clock properties of HCI clock candidates.} The projected instability $\sigma_{\tau}$, differential magnetic- ($\Delta\alpha_{M1}$) and electric-dipole ($\Delta\alpha_{E1}$) polarizability of over 700 HCI clock candidates are plotted, with the grey lines and crosses being their projections in the $x-y$ plane. Each line corresponds to a relevant isoelectronic sequence denoted by the colored cells in the periodic table Fig.~\ref{pd}. 
The values for state-of-the-art singly charged ion clocks~\cite{ludlow2015optical,BBR2018}, Ar$^{13+}$~\cite{HCIclock-Ar13-2022}, Pr$^{9+}$~\cite{HCIclock-4f5p-2019}, Cf$^{15+}$~\cite{HCIclock-5f6p-2012}, and nuclear clock $^{229}$Th$^{3+}$~\cite{Th-2023-Beloy} are presented for comparison.} 
\end{figure}         

In total, we identified more than 700 HCI clock candidates. Their key properties and their comparison with state-of-the-art clocks are presented in Fig.~\ref{clock}. Here, each line in the figure corresponds to one of the 24 isoelectronic sequences, ranging from $2p_{{}^+}^2$ up to $5d_{{}^+}^4$ (the $5f_{{}^\pm}^m$ ions are not included as they are all from elements with a nuclear charge $Z>92$). The projected instability characterizes statistical uncertainty (or precision) of the clock candidates, and the differential electric- and magnetic-dipole polarizabilities characterize the systematic shifts (or accuracies) under the perturbation of external electromagnetic fields. Details can be found in the supplementary Figure~S3. In comparison to other constructed and planned clock references~\cite{ludlow2015optical,kozlov2018highly}, as depicted by the crosses in Fig.~\ref{clock}, these HCI clock references are far less sensitive to perturbations of external electromagnetic fields. Therefore, they would result in black-body radiation shifts, AC Stark shifts and second-order Zeeman shifts, all up to 3 to 5 orders of magnitude smaller than those of state-of-the-art clocks~\cite{rosenband2008frequency,Al+-clock2019}, as well as the planned $^{229}$Th$^{3+}$ nuclear clock~\cite{peik2003nuclear}, rendering it possible to demonstrate clocks with fractional precision $\delta\nu/\nu<10^{-20}$. High-precision spectroscopy~\cite{schmidt2005spectroscopy} of the isotope shifts of this vast range of narrow transitions would find applications in, e.g., King-plot analysis~\cite{5force2018,Yb+5force2020,Ca+5force2020}, and thus enriching the test of nuclear theories and the search for physics beyond the standard model~\cite{newphysics-RMP2018}.

\begin{figure*}[t]
\includegraphics[width=0.85\textwidth]{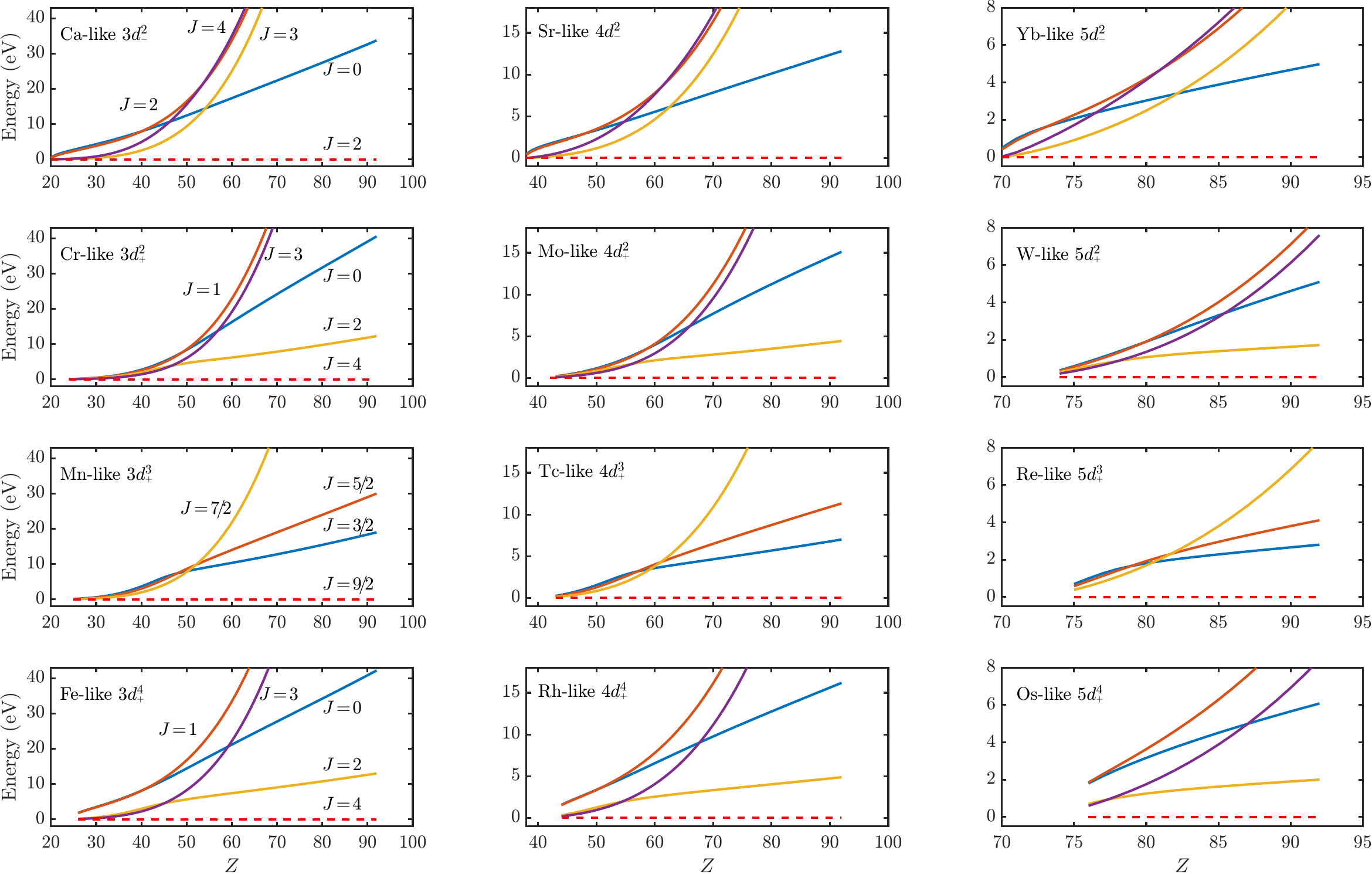}
\caption{\label{HCInd}\textbf{Universal linear scaling laws manifested by mutual-electron interactions.} Energies of the low lying levels of the $nd_{{}^-}^2$, $nd_{{}^+}^2$, $nd_{{}^+}^3$, and $nd_{{}^+}^4$ ions are plotted as functions of the atomic number $Z$. The energies scale linearly with $Z$ are all from the ground-state multiplets listed at the bottom of the periodic table in Fig.~\ref{pd}. } 
\end{figure*}

\textit{Coulomb splitting} ---  
The emergence of the physics discovered above can be illustrated in Fig.~\ref{HCInd}. Here, the evolution of the energies of the low-lying states, as a function of nuclear charge $Z$, are plotted for isoelectronic systems of $nd_{{}^-}^2$, $nd_{{}^+}^2$, $nd_{{}^+}^3$, and $nd_{{}^+}^4$ ($n=3,4,5$) configurations. While these states intertwined together 
in the cases of light elements (i.e., small $Z$), it is obvious that their excitation energies evolve into two distinct scaling laws for heavy ions (i.e., large $Z$). For states involving electron excitations between the $l_{{}^-}$ and $l_{{}^+}$ orbitals, their energies scale as $Z^4/n^3$ that feature the $Z^2Z^{*2}/n^3$ scaling laws for the relativistic fine-structure splittings~\cite{grant2007relativistic,johnson2007atomic} ($Z^*$ is the effective charge under electron screening effect).

For states whose energies scale linearly with $Z$ in Fig.~\ref{HCInd}, they are all from the corresponding ground-state configurations listed in the bottom row of the periodic table (Fig.~\ref{pd}). In the case of $nd_{{}^+}^2$, our analytical calculation predicts 
\begin{eqnarray} 
 E_2-E_4&=&\frac{24}{175}R_2-\frac{2}{63}R_4,\label{e24}\\
 E_0-E_4&=&\frac{12}{35}R_2+\frac{1}{9}R_4.\label{e02} 
\end{eqnarray}
Here, $E_{0,2,4}$ are the energies of the $J=0,2,4$ states, respectively, with $R_k=\int dr_1\int dr_2\rho(r_1)\rho(r_2)r^{k}_{<}/r^{k+1}_{>}$ [$k=2,4$, and $r_>=\max(r_1,r_2)$ and $r_<=\min(r_1,r_2)$] being the Slater integral of rank $k$~\cite{grant2007relativistic,johnson2007atomic}, and $\rho(r)$ the electron density of the radial wavefunction of the $nd_{{}^+}$ orbital. As $R_k\propto\left\langle\frac{1}{r}\right\rangle\propto Z/n^2$~\cite{grant2007relativistic} is a measure of the electron-electron interactions, Eqs.~(\ref{e24},\ref{e02}) indicate that there is a significant cross cancellation of relativistic effects in the energy splittings between these states. Due to such a linear scaling law, as shown in Fig.~\ref{HCInd}, the energies of these states quickly become well isolated from those of the states involving inter-orbital excitations. As a consequence, it leads to the reordering of the fine-structure terms, and to the highly forbidden optical and XUV transitions discussed in Fig.~\ref{clock}. The slopes of the lines would thus directly reflect the strengths of the electron-electron interactions in these ions. In the periodic table of Fig.~\ref{pd}, the reference ions at which the first term reordering happens are displayed at the bottom of each cell. This mechanism also explains the grouped level structures of the excited $3d_{{}^-}^{4-q}3d_{{}^+}^{q}$ configurations observed in Fig~\ref{levels}.

In conclusion, our newly introduced periodic table facilitates the spectroscopic analysis of the grouped level structures in complex highly charged atomic systems, which might find applications in various plasma researches. With a single, simple theory under angular momentum couplings of relativistic orbitals, we are able to discover and categorize hundreds of ultrastable HCI clock candidates, whose projected performance could surpass state-of-the-art clocks by several orders of magnitude.  
Furthermore, with both numerical and analytical methods, we also identify the universal $Z/n^2$ scaling laws in the energy splittings between states from the same relativistic configurations in HCIs. 
As these splittings are a consequence of the electron-electron interaction terms in the multielectron atomic Hamiltonian~\cite{grant2007relativistic,johnson2007atomic}, 
the observed linear scaling laws complement the quadratic $Z^2/n^2$ scaling laws in principal energies governed by the central Coulomb potential of the nucleus, and the $Z^4/n^3$ scaling laws in fine-structure splittings manifested by the relativistic effects. To distinguish it from the well-known fine-structure splitting, we suggest to name this type of splitting ``Coulomb splitting''. These $Z$ scaling laws would provide a systematic understanding of the spectroscopic properties of HCIs in various plasma conditions. 



%

~\\
\noindent{\textbf{End Matter}}\\
\textit{Analytical Methods} -- 
To obtain all the total angular momenta $J$ for the ground-state configuration of each cell in the periodic table Fig.~\ref{pd}, we employ the $jj$ coupling scheme based on the particle creation and annihilation operators. For two-electron systems, this is~\cite{johnson2007atomic} 
\begin{eqnarray} 
 |JM\rangle&=&\sum_{m_a,m_b}C_{ab}^{JM}a^{\dagger}_{a}a^{\dagger}_{b}|0\rangle.\label{CG}
\end{eqnarray}
Here, $|JM\rangle$ is the energy eigenstate with a total angular momentum $J$ and magnetic component $M$, and $|0\rangle$ the vacuum state. $a^{\dagger}_{i}$ ($i=a,b$) creates one electron in the relativistic orbital with angular momentum $j_i$ and magnetic component $m_i$. $C_{ab}^{JM}=C(j_a,j_b,J;m_a,m_b,M)$ with ($M=m_a+m_b$) is the corresponding Clebsch-Gordan coefficient that can be calculated via software such as Mathematica. As we are interested in the ground-state configurations, such as the $np_{{}^+}^2$, $nd_{{}^-}^2$, $nd_{{}^+}^2$, $nf_{{}^-}^2$, and $nf_{{}^+}^2$, one always has $j_a=j_b$. After accounting for the Pauli's exclusion principle, all allowed combinations of electron occupations can be determined, and are listed in the supplementary Tables~S1-S4. With similar procedures, one can also obtain the CSFs for three and four electron systems, namely, the $nd_{{}^+}^3$, $nf_{{}^-}^3$, $nf_{{}^+}^3$, and $nf_{{}^+}^4$ configurations. 

In the following, we take the $nd_{{}^+}^2$ and $nf_{{}^-}^2$ configurations as an example to demonstrate the analytical considerations discussed in the main text. As shown in Table~S2, there is only one occupation scheme for both $|M=4\rangle$ and $|M=3\rangle$ state. As these two Fock states will be accounted into the magnetic levels for the state with $J=4$, it means that there is no state with $J=3$ allowed in such a configuration. With similar considerations, one can also exclude the $J=1$ state. Therefore, only the states with $J=4,2,0$ are allowed in the $nd_{{}^+}^2$ and $nf_{{}^-}^2$ configurations. With the help of Eq.~(\ref{CG}), the three $|M=0\rangle$ Fock states can be used to construct the corresponding three atomic states
\begin{eqnarray} 
 |00\rangle&=&\sqrt{\frac{1}{3}}S_{\frac{1}{2},\frac{-1}{2}}-\sqrt{\frac{1}{3}}S_{\frac{3}{2},\frac{-3}{2}}+\sqrt{\frac{1}{3}}S_{\frac{5}{2},\frac{-5}{2}},\nonumber\\
 |20\rangle&=&-\sqrt{\frac{8}{21}}S_{\frac{1}{2},\frac{-1}{2}}+\sqrt{\frac{1}{42}}S_{\frac{3}{2},\frac{-3}{2}}+\sqrt{\frac{25}{42}}S_{\frac{5}{2},\frac{-5}{2}},\nonumber\\
 |40\rangle&=&\sqrt{\frac{2}{7}}S_{\frac{1}{2},\frac{-1}{2}}+\sqrt{\frac{9}{14}}S_{\frac{3}{2},\frac{-3}{2}}+\sqrt{\frac{1}{14}}S_{\frac{5}{2},\frac{-5}{2}},\nonumber
\end{eqnarray}
with the Slater determinants being defined as 
\begin{eqnarray} 
 S_{ab}&=&\frac{1}{\sqrt{2}}\left(a^{\dagger}_{a}a^{\dagger}_{b}-a^{\dagger}_{b}a^{\dagger}_{a}\right)|0\rangle=\frac{1}{\sqrt{2}}\left(|ab\rangle-|ba\rangle\right).\nonumber
\end{eqnarray}

With the above wavefunctions, one can proceed to calculate the energy of these atomic states. By defining the closed-shell core as the vacuum state, one has the energy 
 \begin{eqnarray} 
 E_{JM} &=&2\epsilon+\sum_{m_am_b}\sum_{m_cm_d}
C_{ab}^{JM}C_{cd}^{JM}g_{abcd}, \nonumber
 \end{eqnarray}
with $\epsilon$ being the one-particle energy of the electron in the $nd_{{}^+}$ or $nf_{{}^-}$ orbital, and $g_{abcd}=\left\langle ab\right|\frac{1}{r_{12}}\left| cd\right\rangle$ the 2-body matrix elements describing electron-electron Coulomb interactions. With further mathematical derivations, it can be proved that $E_{JM}$ is independent of the magnetic component $M$ such that  
\begin{eqnarray} 
E_4&=&2\epsilon+R_0-\frac{4}{35}R_2-\frac{1}{63}R_4,\\
E_2&=&2\epsilon+R_0+\frac{4}{175}R_2-\frac{1}{21}R_4,\\
E_0&=&2\epsilon+R_0+\frac{8}{35}R_2+\frac{2}{21}R_4.
\end{eqnarray}
Here, $R_k=\int dr_1\int dr_2\rho(r_1)\rho(r_2)r^{k}_{<}/r^{k+1}_{>}$ [$k=0,2,4$, and $r_>=\max(r_1,r_2)$ and $r_<=\min(r_1,r_2)$] is the Slater integral of rank $k$~\cite{grant2007relativistic}, and $\rho(r)$ the radial electron density of the $nd_{{}^+}$ or $nf_{{}^-}$ orbital. 
As $R_k$ is always positive and $R_2> R_4$, it is obvious that $E_4<E_2<E_0$. The common term $2\epsilon+R_0$ in the energies of all three CSFs indicates that the excitation energies of the $J=2,0$ states are mainly determined by electron-electron interactions. 

~\\
\textit{Numerical Methods} --- 
The level structures, lifetimes and line strengths mentioned in the main text were calculated via the \textit{ab initio} fully relativistic multiconfiguration Dirac--Hartree--Fock (MCDHF) and relativistic configuration interaction (RCI) methods implemented in the GRASP2018 code~\cite{fischer2019grasp2018}. Within this approach, the many-electron atomic state function (ASF)
is constructed as a linear combination of configuration state functions (CSFs) with common total angular momentum ($J$), magnetic ($M$), and parity ($P$) quantum \mbox{numbers:} 
\[|\Gamma P J M\rangle = \sum_{k} c_k |\gamma_k P J M\rangle.\]
Each CSF $|\gamma_k P J M\rangle$ is built from products of one-electron orbitals (Slater
determinants), $jj$-coupled to the appropriate angular symmetry and parity, and $\gamma_k$ represents orbital occupations, together with orbital and intermediate
quantum numbers necessary to uniquely define the CSF. $\Gamma$ collectively denotes \mbox{all} the $\gamma_k$ involved in the representation of the ASF. $c_k$ is the corresponding mixing coefficient. The radial wave function for each orbital is obtained via solving the self-consistent MCDHF equations under the Dirac--Coulomb Hamiltonian. Then, the RCI method is applied to account for corrections arising from mass shift, quantum electrodynamic and Breit interactions. 

As there are more than 700 HCI clock candidates, and to identify the positions of the reference ions listed in the periodic table Fig.~\ref{pd}, more than 1000 ions need to be calculated. Therefore, it is time consuming to fully account for the correlation effects for all these ions. Luckily, the correlation effects may shift the excitation energies, they do not modify the conclusions discussed in the main text. Therefore, only the correlation effects arising from single-electron excitations from the \{$n(l-1)$, $nl$\} valence orbitals to all the \{$n+1$\} correlation orbitals are included in the GRASP2018 calculations. 



\begin{acknowledgments}
 C.L thanks Dr. Li Han and Dr. Suvam Singh for for discussions and reading the manuscript.
\end{acknowledgments}

\setcounter{figure}{0}
\begin{figure*}[h]
\renewcommand{\figurename}{Figure}
\renewcommand{\thefigure}{S\arabic{figure}}
\includegraphics[width=0.95\textwidth]{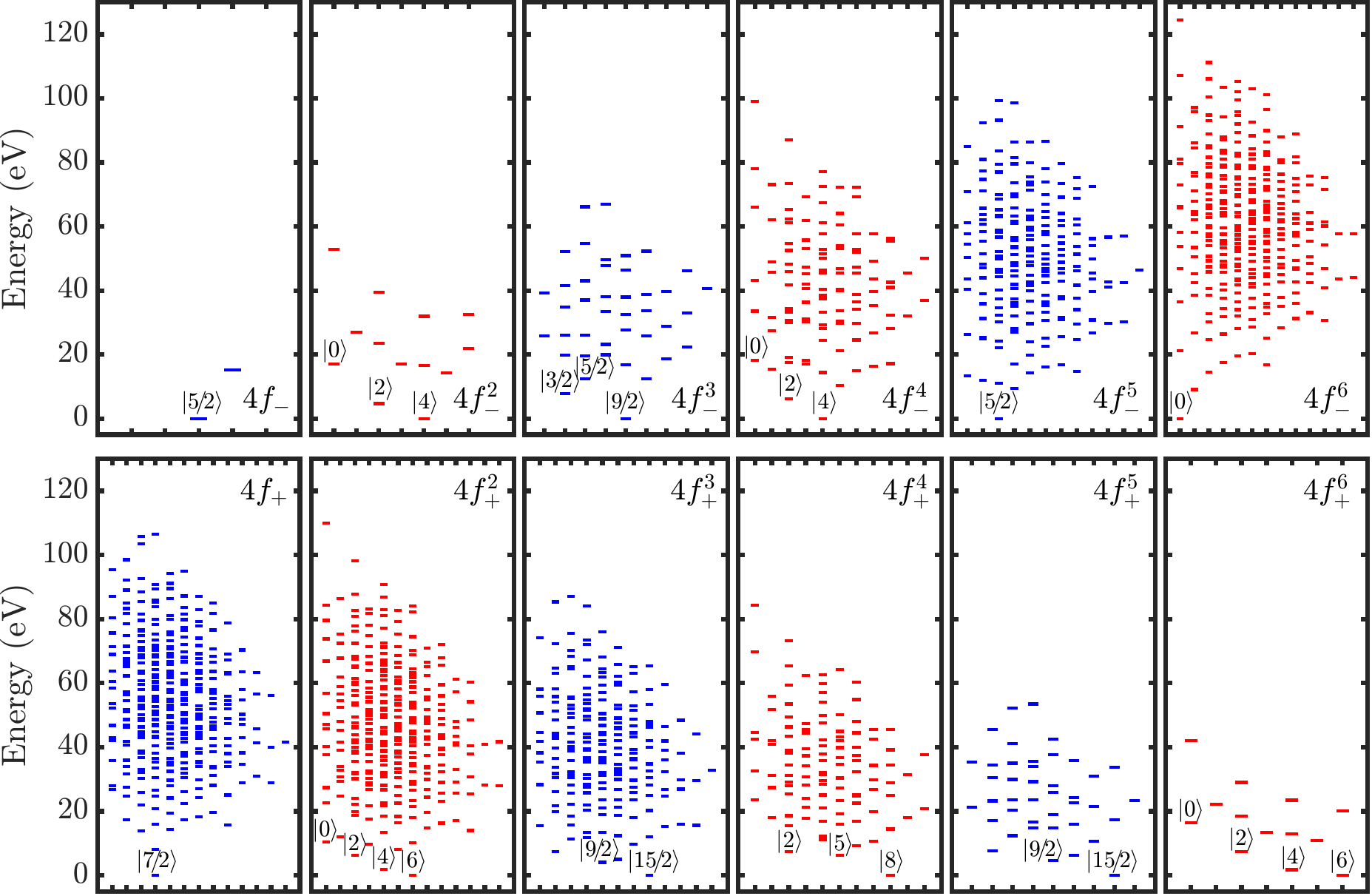} 
\caption{\label{levels4f}\textbf{Level structure of highly charged uranium ions:} the seventh row of the periodic table shown in Fig.~\ref{pd}.} 
\end{figure*}

\begin{figure*}[h]
\renewcommand{\figurename}{Figure}
\renewcommand{\thefigure}{S\arabic{figure}}
\includegraphics[width=0.95\textwidth]{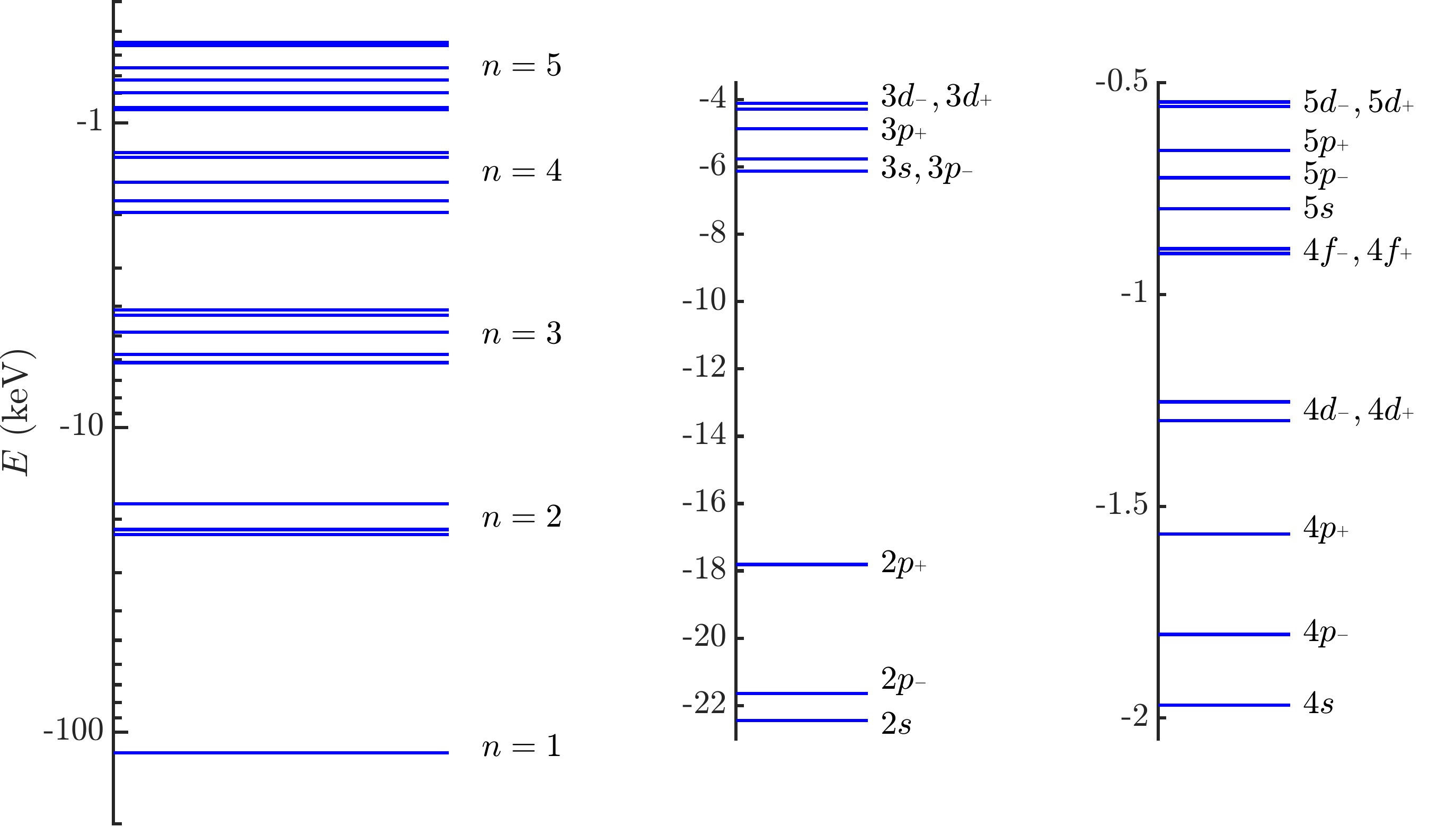}
\caption{\label{levels4f}\textbf{Energies of the relativistic orbitals in U$^{22+}$ ([Er]$5d_{{}^-}^2$).} Results are obtained via DHF calculations.} 
\end{figure*}

\begin{figure*}[b!]
\renewcommand{\figurename}{Figure}
\renewcommand{\thefigure}{S\arabic{figure}}
\includegraphics[width=0.95\textwidth]{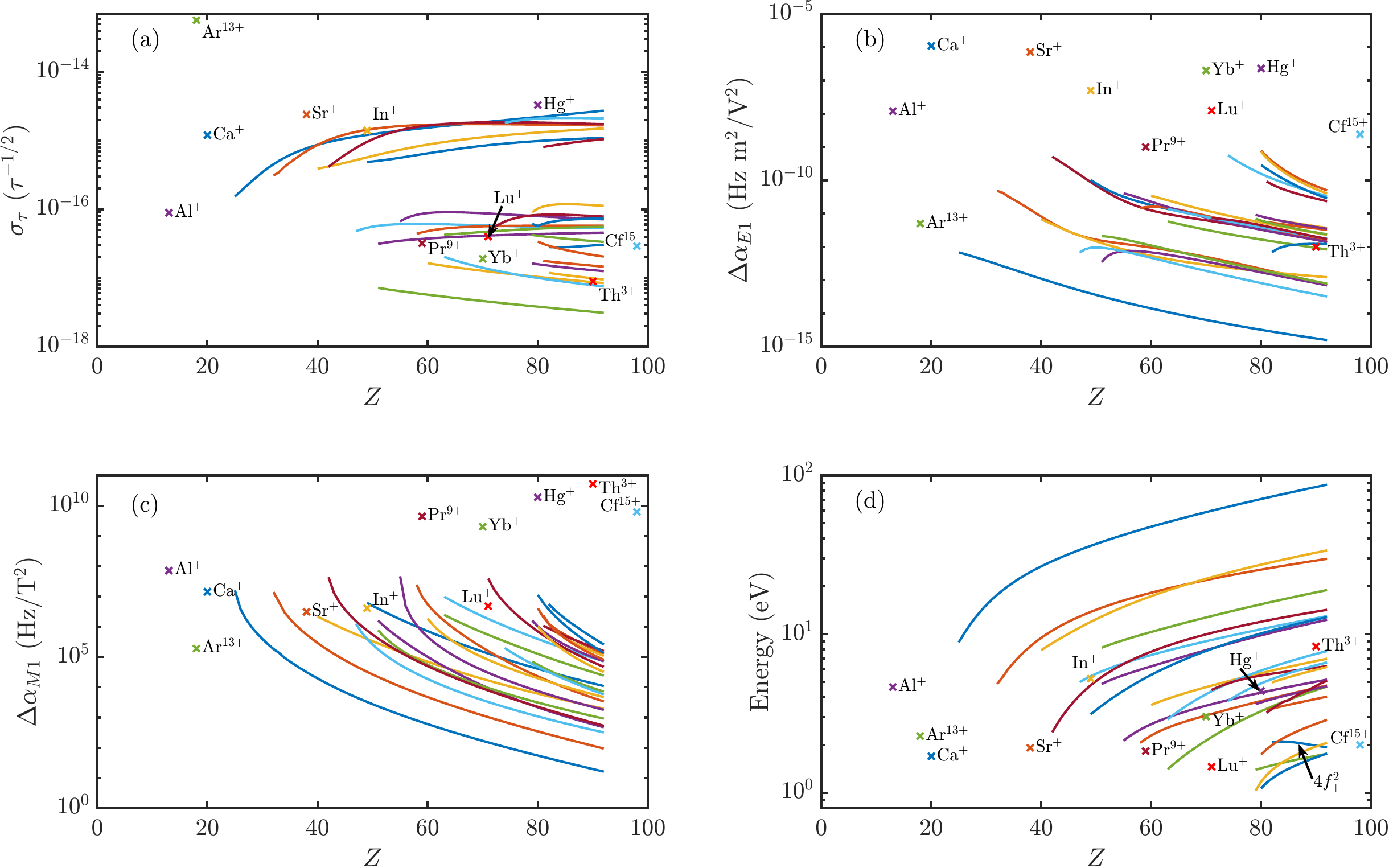} 
\caption{\label{clockall}\textbf{Superior clock properties of the HCI clock candidates.} \textbf{(a)} the projected instability $\sigma_{\tau}$, \textbf{(b)} the differential electric-dipole polarizability $\Delta\alpha_{E1}$, \textbf{(c)} differential magnetic-dipole polarizability $\Delta\alpha_{M1}$, and \textbf{(d)} transition energy of the HCI clock candidates are plotted as a function of the nuclear charge $Z$. With a total of more than 700 HCIs, each line in \textbf{a-c} corresponds to a relevant isoelectronic sequence denoted by the colored cells in the periodic table Fig.~\ref{pd}. For odd $Z$, we have neglected hyperfine interactions which is valid for their even isotopes with a zero nuclear spin. The details of their values can be found in the supplements. The corresponding values for singly charged ion clocks~\cite{ludlow2015optical,BBR2018}, Ar$^{13+}$~\cite{HCIclock-Ar13-2022}, Pr$^{9+}$~\cite{HCIclock-4f5p-2019}, Cf$^{15+}$~\cite{HCIclock-5f6p-2012}, and the nuclear clock $^{229}$Th$^{3+}$~\cite{Th-2023-Beloy} are presented for comparison. For the energies shown in \textbf{(d)}, there is an exceptional line corresponding to the Xe-like isoelectronic sequence ($4f_{^{+}}^2$). This line shows that the transition energy decreases for large $Z$. We found that it is due to a strong mixing between the states in the $4f_{^{+}}^2$ configuration with the states in the $4f_{^{-}}^{6-q}4f_{^{+}}^{2+q}$ configurations. Here, $q$ is the number of electrons being excited from the $4f_{^{-}}$ orbital to the $4f_{^{+}}$ orbital. When only the $4f_{^{+}}^2$ configuration is considered, one would obtain the same increasing trend as observed in the other isoelectronic sequences.}  
\end{figure*}

\clearpage
\begin{samepage}

\begin{table}[h!]
\scriptsize
\renewcommand{\tablename}{Table}
\renewcommand{\thetable}{S\arabic{table}}
\caption{\textbf{Basis set for the Slater determinants of $np_{{}^+}^2$and $nd_{{}^-}^2$ configurations.} In total, there are $C_4^2=6$ states that can be used to construct the $J=2,0$ CSFs.}
\begin{tabular}{c|c|c|c|c|c|c|c|}
  \hline
\backslashbox{$M_J$}{$m_j$}  & 3/2 & 1/2 & -1/2 & -3/2    \\
 \hline
2 & $\times$ & $\times$ &   &     \\
1 & $\times$ &  & $\times$  &      \\
0 & $\times$ &  &   & $\times$    \\
0 &   & $\times$ & $\times$ &      \\
-1 &   & $\times$ &  & $\times$     \\
-2 &   &  & $\times$ & $\times$    
\label{np2}
\end{tabular}
\end{table}

\begin{table}[h!]
\scriptsize
\renewcommand{\tablename}{Table}
\renewcommand{\thetable}{S\arabic{table}}
\caption{\textbf{Basis set for the Slater determinants of the $nd_{{}^+}^2$and $nf_{{}^-}^2$ configurations.} In total, there are $C_6^2=15$ states that can be used to construct the $J=4,2,0$ CSFs.}
\begin{ruledtabular}
\begin{tabular}{c|c|c|c|c|c|c|c|}
  \hline
\backslashbox{$M_J$}{$m_j$} & 5/2 & 3/2 & 1/2 & -1/2 & -3/2  &   -5/2\\
 \hline
4 & $\times$ & $\times$ &   &   &    & \\
3 & $\times$ &  & $\times$  &   &    & \\
2 & $\times$ &  &   & $\times$  &    & \\
1 & $\times$ &  &   &   &  $\times$  & \\
0 & $\times$ &  &   &   &    & $\times$\\
2 &  & $\times$ &  $\times$ &   &    & \\
1 &  & $\times$ &  & $\times$   &    & \\
0 &  & $\times$ &  &    &  $\times$  & \\
-1 &  & $\times$ &  &    &    & $\times$\\
0 &  &  & $\times$ &  $\times$  &    & \\
-1 &  &  & $\times$ &    & $\times$   & \\
-2 &  &  & $\times$ &    &   & $\times$ \\
-2 &  &  &  &  $\times$  & $\times$  &  \\
-3 &  &  &  &  $\times$  &   & $\times$ \\
-4 &  &  &  &    &  $\times$ & $\times$ 
\label{nd2}
\end{tabular}
\end{ruledtabular}
\end{table}

\begin{table}[h!]
\scriptsize
\renewcommand{\tablename}{Table}
\renewcommand{\thetable}{S\arabic{table}}
\caption{Basis set for the Slater determinants of the $nd_{{}^+}^3$and $nf_{{}^-}^3$ configurations. 
In total, there are $C_6^3=20$ states that can be used to construct the $J=9/2,3/2,5/2$ CSFs. 
}
\begin{tabular}{c|c|c|c|c|c|c|c|}
  \hline
\backslashbox{$M_J$}{$m_j$} & 5/2 & 3/2 & 1/2 & -1/2 & -3/2  &   -5/2\\
 \hline
9/2 & $\times$ & $\times$ & $\times$  &   &    & \\
7/2 & $\times$ & $\times$ &   & $\times$  &    & \\
5/2 & $\times$ & $\times$ &   &   &  $\times$  & \\
3/2 & $\times$ & $\times$ &   &   &    & $\times$\\
5/2 & $\times$ &  & $\times$  & $\times$  &    & \\
3/2 & $\times$ &  & $\times$  &   & $\times$   & \\
1/2 & $\times$ &  & $\times$  &   &    & $\times$\\
1/2 & $\times$ &  &   & $\times$  &  $\times$   & \\
-1/2 & $\times$ &  &   & $\times$  &     & $\times$\\
-3/2 & $\times$ &  &   &   &  $\times$  & $\times$\\
3/2 &  & $\times$ &  $\times$ & $\times$ &    & \\
1/2 &  & $\times$ &  $\times$ &  &  $\times$  & \\
-1/2 &  & $\times$ &  $\times$ &  &    & $\times$\\
-1/2 &  & $\times$ &  & $\times$   & $\times$   & \\
-3/2 &  & $\times$ &  & $\times$   &    & $\times$\\
-5/2 &  & $\times$ &  &    &  $\times$  & $\times$\\
-3/2 &  &  & $\times$ &  $\times$  &  $\times$  & \\
-5/2 &  &  & $\times$ &  $\times$  &    & $\times$\\
-7/2 &  &  & $\times$ &    & $\times$   & $\times$\\
-9/2 &  &  &  &  $\times$  & $\times$  & $\times$ 
\label{nd3}
\end{tabular}
\end{table}

\begin{table}[h!]
\scriptsize
\renewcommand{\tablename}{Table}
\renewcommand{\thetable}{S\arabic{table}}
\caption{\textbf{Basis set for the Slater determinants of the $nf_{{}^+}^2$and $ng_{{}^-}^2$ configurations.} In total, there are $C_8^2=28$ states that can be used to construct the $J=6,4,2,0$ CSFs.}
\begin{tabular}{c|c|c|c|c|c|c|c|c|}
  \hline
\backslashbox{$M_J$}{$m_j$} & 7/2 &5/2 & 3/2 & 1/2 & -1/2 & -3/2  &  -5/2 & -7/2\\
 \hline
 6 & $\times$ & $\times$ &   &   &    & &    & \\
 5 & $\times$ &  & $\times$  &   &    & &    & \\
 4 & $\times$ &  &   & $\times$  &    & &    &  \\
 3 & $\times$ &  &   &   &  $\times$  & &    &  \\
 2 & $\times$ &  &   &   &    & $\times$ &    & \\
 1 & $\times$ &  &   &   &    &  &  $\times$  & \\
 0  & $\times$ &  &   &   &    &  &    & $\times$\\
 4 &  & $\times$ &  $\times$ &   &    &  &    & \\
 3 &  & $\times$ &  & $\times$   &    & &    &  \\
 2 &  & $\times$ &  &    &  $\times$  &  &    & \\
 1 &  & $\times$ &  &    &    & $\times$ &    & \\
 0  &  & $\times$ &  &    &    &  &  $\times$  & \\
 -1  &  & $\times$ &  &    &    &  &    & $\times$\\
 2 &  &  & $\times$ &  $\times$  &    &  &    &\\
 1 &  &  & $\times$ &    & $\times$   &  &    &\\
 0  &  &  & $\times$ &    &   & $\times$  &    &\\
 -1  &  &  & $\times$ &    &   &  &   $\times$  &\\
 -2  &  &  & $\times$ &    &   &   &    &$\times$\\
 0  &  &  &  &  $\times$  & $\times$  &   &    &\\
 -1  &  &  &  &  $\times$  &   & $\times$  &    &\\
 -2  &  &  &  &  $\times$  &   &   &  $\times$  &\\
 -3  &  &  &  &  $\times$  &   &   &    & $\times$\\
 -2  &  &  &  &    &  $\times$ & $\times$  &    & \\
 -3  &  &  &  &    &  $\times$ &   &  $\times$  & \\
 -4  &  &  &  &    &  $\times$ &  &    &  $\times$\\
 -4  &  &  &  &    &   & $\times$  &  $\times$  & \\
 -5  &  &  &  &    &   & $\times$  &    & $\times$\\
 -6  &  &  &  &    &   &  &  $\times$  &  $\times$
\label{nf2}
\end{tabular}
\end{table}

\end{samepage}

\end{document}